\documentclass[twocolumn,pre,aps,showpacs,showkeys,amsmath]{revtex4}
\usepackage{graphicx}
\usepackage{adjustbox}
\usepackage{multirow}

\begin{document}

\title{Break-up and Recovery of Harmony between Direct and Indirect Pathways in The Basal Ganglia; Huntington's Disease and Treatment}
\author{Sang-Yoon Kim}
\email{sykim@icn.re.kr}
\author{Woochang Lim}
\email{wclim@icn.re.kr}
\affiliation{Institute for Computational Neuroscience and Department of Science Education, Daegu National University of Education, Daegu 42411, Korea}

\begin{abstract}
The basal ganglia (BG) in the brain exhibit diverse functions for motor, cognition, and emotion.
Such BG functions could be made via competitive harmony between the two competing pathways, direct pathway (DP) (facilitating movement) and indirect pathway (IP) (suppressing movement). As a result of break-up of harmony between DP and IP, there appear pathological states with disorder for movement, cognition, and psychiatry. In this paper, we are concerned about the Huntington's disease (HD), which is a genetic neurodegenerative disorder causing involuntary movement and severe cognitive and psychiatric symptoms. For the HD, the number of D2 SPNs ($N_{\rm D2}$) is decreased due to degenerative loss, and hence, by decreasing $x_{\rm D2}$ (fraction of $N_{\rm D2}$), we investigate break-up of harmony between DP and IP in terms of their competition degree ${\cal C}_d$, given by the ratio of strength of DP (${\cal S}_{DP}$) to strength of IP (${\cal S}_{IP}$) (i.e., ${\cal C}_d = {\cal S}_{DP} / {\cal S}_{IP}$). In the case of HD, the IP is under-active, in contrast to the case of Parkinson's disease with over-active IP, which results in increase in ${\cal C}_d$ (from the normal value).
Thus, hyperkinetic dyskinesia such as chorea (involuntary jerky movement) occurs. 
We also investigate treatment of HD, based on optogenetics and GP ablation, by increasing strength of IP, resulting in recovery of harmony between DP and IP. Finally, we study effect of loss of healthy synapses of all the BG cells on HD. Due to loss of healthy synapses, disharmony between DP and IP increases, leading to worsen symptoms of the HD.
\end{abstract}


\maketitle

\section{Introduction}
\label{sec:INT}
The basal ganglia (BG) (called the dark basement of the brain) are a group of subcortical deep-lying nuclei, receiving excitatory cortical input from most areas of cortex, and they provide inhibitory output to the thalamus and brainstem \cite{Luo,Kandel,Squire,Bear}.
The BG exhibit a variety of functions for motor control and regulation of cognitive and emotional processes \cite{Luo,Kandel,Squire,Bear,GPR1,GPR2,Hump1,Hump2,Hump3,Man}.
Dysfunction in the BG is related to movement disorder [e.g., Parkinson's disease (PD) and Huntington's disease (HD)] and cognitive
and psychiatric disorders \cite{Luo,Kandel,Squire,Bear}.

In this paper, we are concerned about the HD. It is a rare hereditary neurodegenerative disease with severe symptoms for motor, cognition, and
emotion \cite{HD1,HD2,HD3,HD4,HD5,HD6,HD7}. As is well known, patients with HD show hyperkinetic dyskinesia such as chorea (involuntary jerky dance-like movement) as well as cognitive (e.g., dementia) and psychiatric (e.g,, depression and anxiety) disorders. In contrast, patients with PD show hypokinetic disorder such as
slowed movement (bradykinesia) \cite{PD1,PD2,PD3,PD4,PD5,PD6}. Thus, if PD lies at one end of the spectrum of movement disorders in the BG, HD lies at the other end.
We note that HD is caused by a mutated huntingtin (HTT) gene on chromosome 4 \cite{HTT1,HTT2}.
As a result of mutation in HTT gene, the defective HTT gene has abnormal excessive repeats of a three-base (CAG) DNA sequence;
in the mutant gene, the repeat occurs over and over again, from 40 times to more than 80.
The greater the number of CAG repeats, the earlier the onset and severity of HD.
This kind of trinucleotide repeat expansion results in production of abnormal HTT protein that accumulates, resulting in creation of toxic HTT protein aggregates damaging neurons (e.g., death of striatal cells in the BG).
Thus, the primary pathological feature of HD is appearance of toxic HTT protein aggregates, causing the characteristic neurodegeneration seen in HD,
in contrast to the case of PD where dopamine (DA) deficiency is a major cause.

In our recent work for the PD in the BG \cite{KimPD}, we made refinements on the spiking neural network (SNN) for the BG, based on the SNNs for the BG developed in previous works \cite{SPN1,SPN2,CN6}; details on the SNN are given in Sec.~II and Appendices in \cite{KimPD}.
This SNN for the BG is based on anatomical and physiological data derived from rat-based works as follows. For the architecture of the BG SNN (e.g., number of BG cells and synaptic connection probabilities), we refer to the anatomical works \cite{Ana4,Ana5,Ana6,Ana7}. For the intrinsic parameter values of single BG neurons, refer to the physiological properties of the BG neurons \cite{Phys1,Phys2,Phys3,Phys4,Phys5,Phys6,Phys7,Phys8,Phys9,Phys10,Phys11}.
For the synaptic parameters (related to synaptic currents), we also refer to the physiological works \cite{Phys12,Phys13,Phys14,Phys15,Phys16,Phys17,Phys18,Phys19,Phys20}.
Here, we use the rat-brain terminology throughout.
The BG receive excitatory cortical input from most regions of cortex via the input nuclei [striatum and subthalamic nucleus (STN)] and project inhibitory output via the output nucleus [substantia nigra pars reticulata (SNr)], through the thalamus to the motor area of the cortex \cite{Hump1,Man}.
We also note that, the principal input nucleus, striatum, is the primary recipient of DA, arising from the substantia nigra pars compacta (SNc).
Within the striatum, spine projection neurons (SPNs), comprising up to 95 $\%$ of the whole striatal population, are the only primary output neurons \cite{Str1,Str2}. There are two types of SPNs with D1 and D2 receptors for the DA. The DA modulates firing activity of the D1 and D2 SPNs in a different way \cite{SPN1,SPN2,CN6}. In the early stage of HD, degenerative loss of D2 SPNs occurs due to mutation in the HTT gene, while DA level in the striatum is nearly normal \cite{Degen1,Degen2,Degen3,Degen4}.

There are two competing pathways, direct pathway (DP) and indirect pathway (IP), in the BG \cite{DIP1,DIP2,DIP3,DIP4}.
D1 SPNs in the striatum make direct inhibitory projection to the output nucleus, SNr, through DP, and then the thalamus becomes disinhibited.
Consequently, movement facilitation occurs. In contrast, D2 SPNs are connected to the SNr through IP, crossing the intermediate
control nucleus, GP (globus pallidus), and the STN. In the case of IP, the firing activity of the SNr becomes enhanced mainly because of excitatory input from the STN. As a result, firing activity of the thalamus becomes decreased, resulting in movement suppression.

Diverse functions of the BG could be made via ``balance'' of DP and IP.
So far, a variety of subjects for the BG have been investigated in many computational works. Diverse neuron models were employed in the computational works;
(a)	artificial neurons of the leaky-integrator type \cite{GPR1,GPR2,Hump3},
(b)	point neuron function using rate-coded output activation \cite{Frank1,Frank2,CN13},
(c)	leaky integrate-and-fire model \cite{Hump1,Hump2},
(d)	adaptive exponential integrate and fire model \cite{CN14,CN15},
(e)	oscillatory model for local field potentials \cite{CN21},
(f)	dendrite model \cite{CN11},
(g)	firing rate model \cite{CNHD3},
(h)	multiple compartments model \cite{CNHD2},
(i)	Hodgkin-Huxley type neuron model \cite{CN20,CNYu1,CNYu3}, and
(j) Izhikevich neuron model \cite{SPN1,Str2,CN16,CN19,CN7,CN9,SPN2,Man,CN2,CN3,CN4,CN10,CN18,CN6,CN8,PD4,CN17,CN5,CNHD1,CN22,CN1,CNYu2}.

But, no quantitative analysis for balance between DP and IP was made.
As a first time, in our recent work \cite{KimPD}, we made quantitative analysis for competitive harmony (i.e., competition and cooperative interplay) between DP and IP by introducing their competition degree ${\cal C}_d$, given by the ratio of strength of DP (${\cal S}_{DP}$) to strength of IP (${\cal S}_{IP}$) (i.e., ${\cal C}_d = {\cal S}_{DP} / {\cal S}_{IP}$); ${\cal S}_{DP}$ $({\cal S}_{IP})$ is given by the magnitude of the total time-averaged synaptic current into the output nucleus, SNr, through DP (IP).

In this paper, we take into consideration of degenerative loss of D2 SPNs for the HD; $N_{\rm D2}$ (number of D2 SPNs) = $N_{\rm D2}^*$ (normal value) $x_{\rm D2}$ [$1 > x_{\rm D2}$ (fraction of number of D2 SPNs) $\geq 0$] \cite{Degen1,Degen2,Degen3,Degen4,Degen5,Degen6}.
By decreasing $x_{\rm D2}$ from 1, we investigate break-up of harmony between DP and IP for the HD by
employing the competition degree ${\cal C}_d$ in the case of normal DA level ($\phi=0.3$). Due to degenerative loss of D2 SPNs, IP becomes under-active (i.e., weakened), leading to increase in ${\cal C}_d$ from normal value. Thus, hyperkinetic dyskinesia such as chorea occurs, which is in contrast to the case of PD
with reduced ${\cal C}_d$, causing hypokinetic disorder.
Next, based on optogenetics \cite{OG1,OG2}, treatment of HD is also studied via recovery of harmony between DP and IP.
Through activation of D2 SPN (STN) and deactivation of GP, IP becomes strengthened and thus harmony between DP and IP may be recovered.
Finally, we investigate effect of loss of healthy synapses of all the BG cells on HD \cite{Loss1,Loss2,Loss3,Loss4,Loss5,Loss6}.

This paper is organized as follows. In the main Sec.~\ref{sec:QA}, in the SNN for the BG \cite{KimPD}, we make quantitative analysis of break-up and recovery of harmony between DP and IP for the HD. In the Supplementary Information (SI), brief description on the SNN for the BG is given.
Finally, we give summary and discussion in Sec.~\ref{sec:SUM}.

\begin{figure}
\includegraphics[width=1.0\columnwidth]{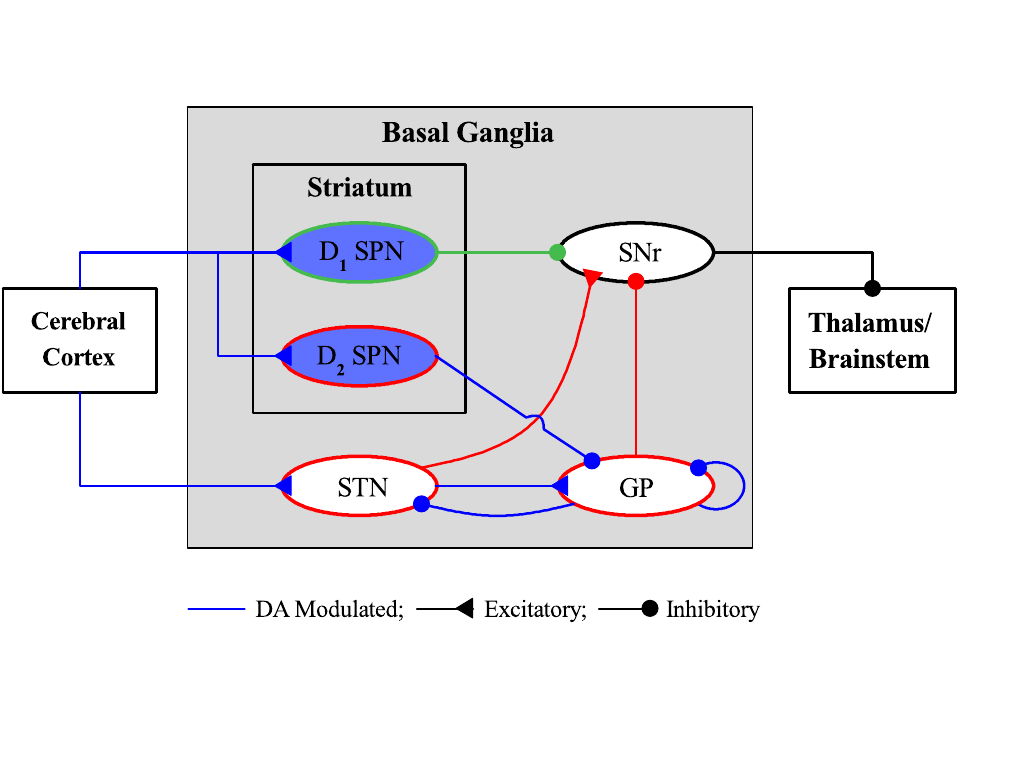}
\caption{Box diagram of our spiking neural network for the basal ganglia (BG). Excitatory and inhibitory connections are denoted by lines with triangles and circles, respectively, and dopamine-modulated cells and connections are represented in blue color. Striatum and STN (subthalamic nucleus), receiving the excitatory cortical input, are two input nuclei to the BG. In the striatum, there are two kinds of inhibitory spine projection neurons (SPNs); SPNs with the D1 receptors (D1 SPNs) and SPNs with D2 receptors (D2 SPNs). The D1 SPNs make direct inhibitory projection to the output nuclei SNr (substantia nigra pars reticulate) through the direct pathway (DP; green color). In contrast, the D2 SPNs are connected to the SNr through the indirect pathway (IP; red color) crossing the GP (globus pallidus) and the STN. The inhibitory output from the SNr to the thalamus/brainstem is controlled through competition between the DP and IP.
}
\label{fig:BGN}
\end{figure}

\section{Quantitative Analysis of Break-up and Recovery of Harmony between DP and IP for the HD}
\label{sec:QA}
In this section, in the SNN for the BG considered in our prior work \cite{KimPD}, we quantitatively analyze competitive harmony (i.e., competition and cooperative interplay) between DP and IP for the HD in terms of the competition degree ${\cal C}_d$ between them, introduced in our previous work \cite{KimPD}. ${\cal C}_d$ is given by the ratio of strength of DP (${\cal S}_{DP}$) to strength of IP (${\cal S}_{IP}$) (i.e., ${\cal C}_d = {\cal S}_{DP} / {\cal S}_{IP}$).

\begin{figure*}
\includegraphics[width=1.6\columnwidth]{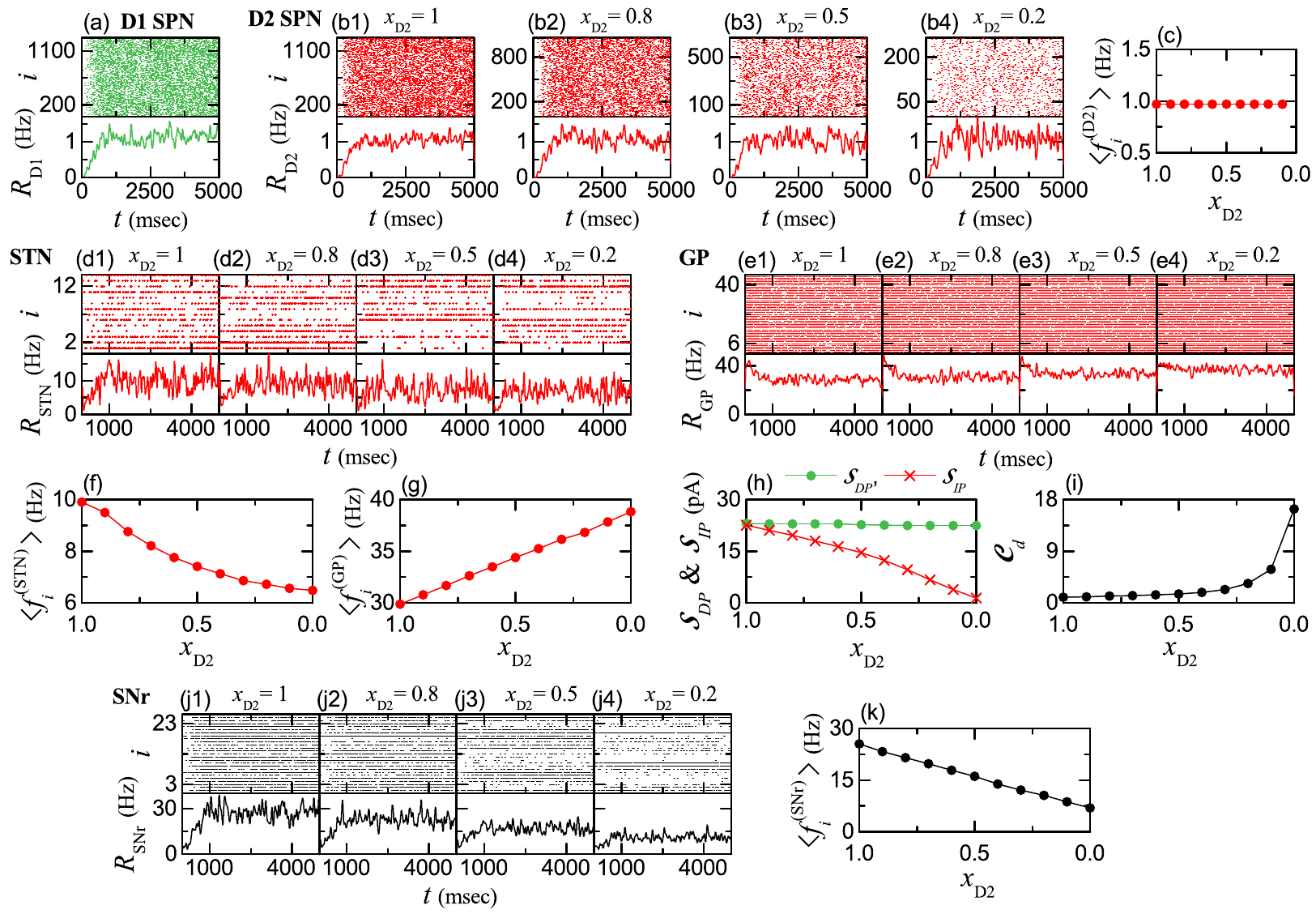}
\caption{Involuntary jerky movement due to degenerative loss of D2 SPNs in the tonic pathological state for the tonic cortical input (3 Hz) in the resting state.
Colors: parts, related to DP (green), while parts, associated with IP (red). (a) Raster plot of spikes and IPSR (instantaneous population spike rate) $R_{\rm D1} (t)$ of D1 SPNs. Raster plots of spikes and IPSRs $R_{\rm D2} (t)$ of D2 SPNs for (b1) $x_{\rm D2}$ = 1.0, (b2) 0.8, (b3) 0.5, and (b4) 0.2. (c) Plot of population-averaged MFR (mean firing rate) $\langle f_i^{({\rm D2})} \rangle$ of D2 SPNs versus $x_{\rm D2}$. Raster plots of spikes and IPSRs $R_{\rm STN} (t)$ of STN neurons for (d1) $x_{\rm D2}$ = 1.0, (d2) 0.8, (d3) 0.5, and (d4) 0.2. Raster plots of spikes and IPSRs $R_{\rm GP} (t)$ of GP neurons for (e1) $x_{\rm D2}$ = 1.0, (e2) 0.8, (e3) 0.5, and (e4) 0.2. Plots of population-averaged MFRs of (f) STN neurons $\langle f_i^{({\rm STN})} \rangle$ and (g) GP neurons $\langle f_i^{({\rm GP})} \rangle$ versus $x_{\rm D2}$. (h) Plots of strengths of DP ${\cal{S}}_{DP}$ and IP ${\cal{S}}_{IP}$ versus $x_{\rm D2}$. (i) Plot of the competition degree ${\cal{C}}_d$ versus $x_{\rm D2}$. Raster plot of spikes and IPSR $R_{\rm SNr} (t)$ of SNr neurons for (j1) $x_{\rm D2}$ =1, (j2) 0.8, (j3) 0.5, and (j4) 0.2. (k) Plot of population-averaged MFR $\langle f_i^{({\rm SNr})} \rangle$ of SNr neurons versus $x_{\rm D2}$.
}
\label{fig:Tonic}
\end{figure*}

As explained in the section of Introduction, we consider the SNN for the BG, based on anatomical and physiological data derived from rat-based studies. 
We note that, in the rat brain, the SNr is the only output nucleus of the BG, in contrast to the higher animals such as humans where both GPi
(internal globus pallidus) and SNr are output nuclei. Figure \ref{fig:BGN} shows a box diagram for the SNN for the BG, consisting of D1/D2 SPNs, STN cells, GP cells, and SNr cells. As the single neuron model of the BG cells, we use the Izhikevich spiking neuron model (which is not only biologically plausible, but also computationally efficient) \cite{Izhi1,Izhi2,Izhi4,Izhi3}. Blue colored cells and lines denote BG cells and synaptic connections affected by the DA. Both striatum and STN receive cortical inputs from most areas of the cortex. We model cortical inputs in terms of 1,000 independent Poisson spike trains with the same firing rate $f$. There are two pathways, DP (green) and IP (red). Inhibitory projection from the D1 SPNs to the output nucleus SNr is provided through the DP. In contrast, D2 SPNs are indirectly connected to the SNr through the IP, crossing the GP and the STN. Inhibitory output from the SNr to the thalamus/brainstem is controlled through competitive harmony between DP and IP \cite{KimPD}. In the SI, brief description on the SNN for the BG is given; for more details, refer to Sec.~II in \cite{KimPD}.

Here, we consider the early stage of HD where neurodegerative loss of D2 SPNs occurs;
$N_{\rm D2}$ (number of D2 SPNs) = $N_{\rm D2}^*$ (=1325; normal value) $x_{\rm D2}$ [$1 > x_{\rm D2}$ (fraction of number of D2 SPNs) $\geq 0$] \cite{Degen1,Degen2,Degen3,Degen4,Degen5,Degen6}. By decreasing $x_{\rm D2}$ from 1, we investigate break-up of harmony between DP and IP in both cases of tonic cortical input (3 Hz) in the resting state and phasic cortical input (10 Hz) in the phasically active state. In these cases, the IP becomes weakened, and thus  ${\cal C}_d$ becomes larger than normal ones. Consequently, involuntary jerky movement and abnormal hyperkinetic movement occur in the tonic and phasic cases, respectively. Next, we study treatment of HD through recovery of harmony between DP and IP. We strengthen the IP via activation of D2 SPNs and STN neurons and deactivation of GP neurons, based on optogenetics \cite{OG1,OG2}. Consequently, harmony between DP and IP becomes recovered, leading to normal movement.
Finally, we also investigate the effect of loss of healthy synapses in the BG neurons on the HD.

\subsection{Break-up of Harmony between DP and IP for the HD}
\label{subsec:HD}
In the early stage of HD, we consider the case of normal DA level of $\phi_1 = \phi_2= \phi= 0.3$ for the D1 and D2 SPNs.
As explained above, cortical inputs are modeled in terms of 1,000 independent Poisson spike trains with firing rate $f$.
We first consider the case of tonic cortical input with $f=3$ Hz in the resting state \cite{Hump1,CI1,CI2,CI3,CI4,CI5,Str2,CN6,CN14}.

Population firing behavior of BG neurons could be well visualized in the raster plot of spikes, corresponding to a collection of spike trains of individual BG neurons. Figure \ref{fig:Tonic}(a) shows the raster plot of spikes for D1 SPNs, associated with DP (green color).
In contrast to the case of D1 SPNs, degenerative loss of D2 SPNs occurs. With decreasing $x_{\rm D2}$ (i.e., fraction of number of D2 SPNs) from 1,
we also get the raster plots of spikes of D2 SPNs [Figs.~\ref{fig:Tonic}(b1)-\ref{fig:Tonic}(b4)], the STN neurons [Figs.~\ref{fig:Tonic}(d1)-\ref{fig:Tonic}(d4)], and the GP neurons [Figs.~\ref{fig:Tonic}(e1)-\ref{fig:Tonic}(e4)], related to the IP (red color) for $x_{\rm D2}=$ 1.0, 0.8, 0.5, and 0.2.

As a collective quantity showing population behaviors, we employ an IPSR (instantaneous population spike rate) which may be obtained from the raster plot of spikes \cite{IPSR1,IPSR2,IPSR3,IPSR4,IPSR5}. Each spike in the raster plot is convoluted with a kernel function $K_h(t)$ to get a smooth estimate of IPSR $R_X(t)$   \cite{Kernel}:
\begin{equation}
R_X(t) = \frac{1}{N_X} \sum_{i=1}^{N_X} \sum_{s=1}^{n_i^{(X)}} K_h (t-t_{s,i}^{(X)}).
\label{eq:IPSR}
\end{equation}
Here, $N_X$ is the number of the neurons in the $X$ population, and $t_{s,i}^{(X)}$ and $n_i^{(X)}$ are the $s$th spiking time and the total number of spikes for the $i$th neuron, respectively.
We use a Gaussian kernel function of band width $h$:
\begin{equation}
K_h (t) = \frac{1}{\sqrt{2\pi}h} e^{-t^2 / 2h^2}, ~~~~ -\infty < t < \infty,
\label{eq:Gaussian}
\end{equation}
where the band width $h$ of $K_h(t)$ is 20 msec.
The IPSRs $R_X(t)$ for $X=$ D1 (SPN), D2 (SPN), STN, GP, and SNr are also shown below their respective raster plots of spikes.
Here, the case of $x_{\rm D2}=1$ corresponds to the normal one without degenerative loss of D2 SPNs.
With decreasing $x_{\rm D2}$ from 1, the population firing activities of the D2 SPNs, the STN neurons, and the GP neurons, associated with IP (red), are changed, while that of the D1 SPN, related to DP (green), is unchanged.

We also study the population-averaged mean firing rate (MFR) of the neurons $\langle f_i^{(X)} \rangle$ in the $X$ population [$X=$ D1 (SPN), D2 (SPN), STN, and GP];
$f_i^{(X)}$ is the MFR of the $i$th neuron in the $X$ population, and $\langle \cdots \rangle$ represents a population average over all the neurons.
For the D1 and D2 SPNs, $\langle f_i^{({\rm D1})} \rangle = 1.03$ Hz and $\langle f_i^{({\rm D2})} \rangle = 0.97$ Hz, independently of $x_{\rm D2}$, because there is no change in cortical inputs to the D1/D2 SPNs; see Fig.~\ref{fig:Tonic}(c) for the D2 SPNs. As $x_{\rm D2}$ is decreased from 1, $\langle f_i^{({\rm GP})} \rangle$ of the GP neurons is increased from 29.9 to 38.8 Hz, due to decrease in inhibitory projection from the D2 SPNs, as shown in Fig.~\ref{fig:Tonic}(g). In contrast, because of increased inhibitory projection from the GP, $\langle f_i^{({\rm STN})} \rangle$ of the STN neurons is decreased from 9.9 to 6.5 Hz [see Fig.~\ref{fig:Tonic}(f)].

We note that, there are two types of synaptic currents into the (output) SNr neurons, $I_{DP}$ and $I_{IP}$, via DP (green) and IP (red) in Fig.~\ref{fig:BGN}, respectively. The DP current, $I_{DP}(t),$ is just the (inhibitory) synaptic current from the D1 SPNs to the SNr neurons:
\begin{equation}
  I_{DP}(t) = - I_{syn}^{({\rm SNr,D1})}(t).
\label{eq:DPC}
\end{equation}
There is no change in $I_{DP}(t),$ independently of $x_{\rm D2}$.

The IP current, $I_{IP}(t),$ is composed of the excitatory component, $I_{IP}^{(E)}(t),$ and the inhibitory component, $I_{IP}^{(I)}(t):$
\begin{equation}
  I_{IP}(t) = I_{IP}^{(E)}(t) + I_{IP}^{(I)}(t).
\label{eq:IPC}
\end{equation}
Here, $I_{IP}^{(E)}(t)$ [$I_{IP}^{(I)}(t)$] is just the synaptic current from the STN (GP) to the SNr:
\begin{equation}
  I_{IP}^{(E)}(t) = - I_{syn}^{({\rm SNr,STN})}(t)~~{\rm and}~~  I_{IP}^{(I)}(t) = - I_{syn}^{({\rm SNr,GP})}(t).
\label{IPCEI}
\end{equation}
Unlike the case of $I_{DP}(t),$ with decreasing $x_{\rm D2}$ from 1, $I_{IP}(t)$ becomes decreased due to decrease in $I_{IP}^{(E)}(t)$ and increase
in $| I_{IP}^{(I)}(t) |$ ($| \cdots |$: absolute magnitude).

Firing activity of the (output) SNr neurons is determined through competition between $I_{DP}(t)$ (DP current) and $I_{IP}(t)$ (IP current)
into the SNr. The strengths of DP and IP, ${\cal S}_{DP}$ and ${\cal S}_{IP}$, are given by the magnitudes of their respective time-averaged synaptic currents:
\begin{equation}
  {\cal S}_{DP} = |\overline{I_{DP}(t)}|~~~{\rm and}~~~{\cal S}_{IP} = |\overline{I_{IP}(t)}|,
\label{eq:Strength}
\end{equation}
where the overline denotes the time averaging and $| \cdots |$ represents the absolute magnitude.
Then, the competition degree ${\cal C}_d$ between DP and IP (given by the ratio of ${\cal S}_{DP}$ to ${\cal S}_{IP}$) was introduced in \cite{KimPD}:
\begin{equation}
{\cal C}_d =  \frac {{\cal S}_{DP}} {{\cal S}_{IP}}.
\label{eq:CD}
\end{equation}

\begin{figure*}
\includegraphics[width=1.6\columnwidth]{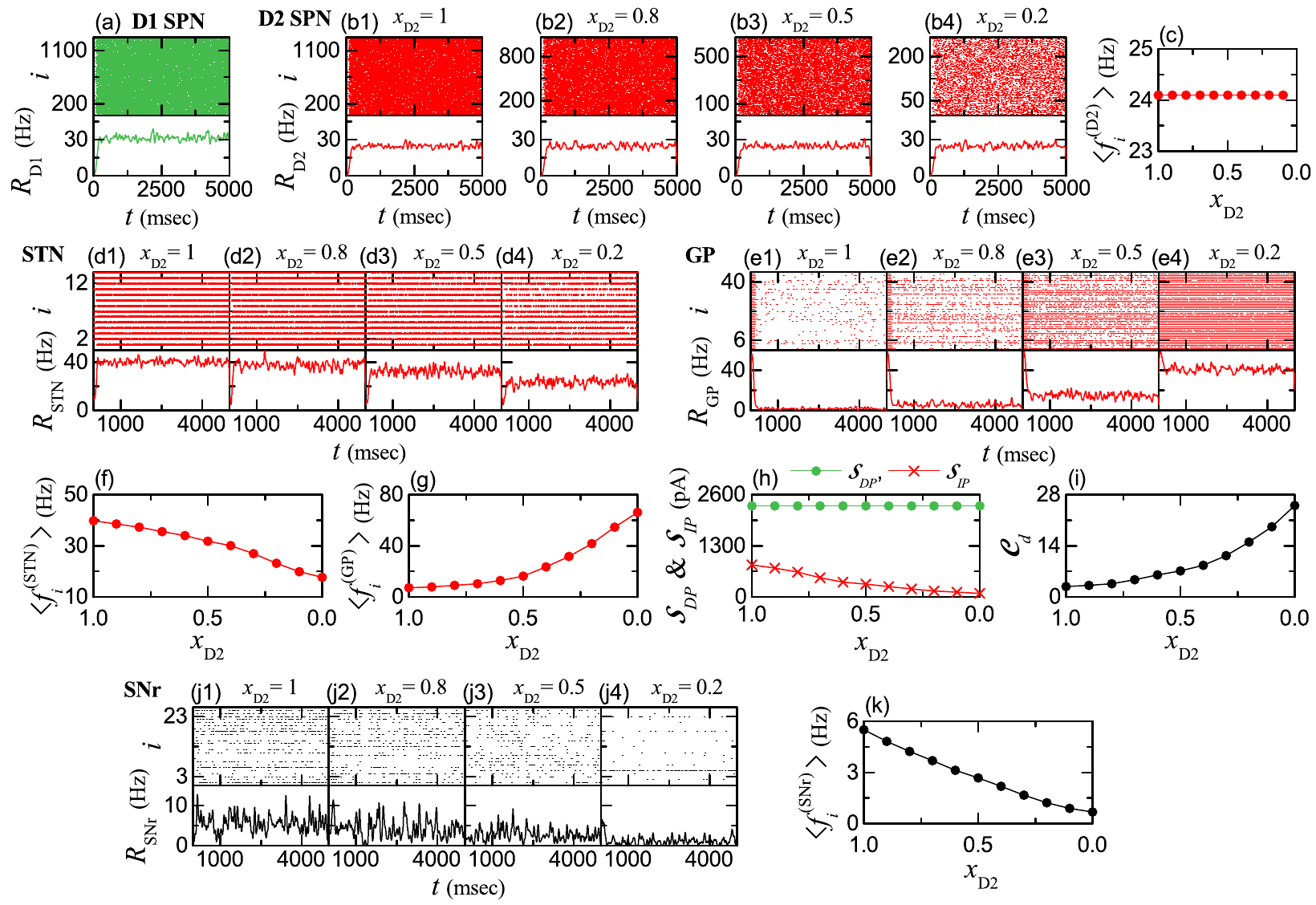}
\caption{Abnormal hyperkinetic movement due to degenerative loss of D2 SPNs in the phasic pathological state for the phasic cortical input (10 Hz) in the phasically-active state. Colors: parts, related to DP (green), while parts, associated with IP (red). (a) Raster plot of spikes and IPSR (instantaneous population spike rate) $R_{\rm D1} (t)$ of D1 SPNs. Raster plots of spikes and IPSRs $R_{\rm D2} (t)$ of D2 SPNs for (b1) $x_{\rm D2}$ = 1.0, (b2) 0.8, (b3) 0.5, and (b4) 0.2. (c) Plot of population-averaged MFR (mean firing rate) $\langle f_i^{({\rm D2})} \rangle$ of D2 SPNs versus $x_{\rm D2}$. Raster plots of spikes and IPSRs $R_{\rm STN} (t)$ of STN neurons for (d1) $x_{\rm D2}$ = 1.0, (d2) 0.8, (d3) 0.5, and (d4) 0.2. Raster plots of spikes and IPSRs $R_{\rm GP} (t)$ of GP neurons for (e1) $x_{\rm D2}$ = 1.0, (e2) 0.8, (e3) 0.5, and (e4) 0.2. Plots of population-averaged MFRs of (f) STN neurons $\langle f_i^{({\rm STN})} \rangle$ and (g) GP neurons $\langle f_i^{({\rm GP})} \rangle$ versus $x_{\rm D2}$. (h) Plots of strengths of DP ${\cal{S}}_{DP}$ and IP ${\cal{S}}_{IP}$ versus $x_{\rm D2}$. (i) Plot of the competition degree ${\cal{C}}_d$ versus $x_{\rm D2}$. Raster plot of spikes and IPSR $R_{\rm SNr} (t)$ of SNr neurons for (j1) $x_{\rm D2}$ =1, (j2) 0.8, (j3) 0.5, and (j4) 0.2. (k) Plot of population-averaged MFR $\langle f_i^{({\rm SNr})} \rangle$ of SNr neurons versus $x_{\rm D2}$.
}
\label{fig:Phasic}
\end{figure*}

For $x_{\rm D2}=1$ (without degenerative loss of D2 SPNs), ${\cal S}_{DP}$ = 23.1 and ${\cal S}_{IP}$ = 23.4, and hence DP and IP become nearly balanced (i.e., ${\cal C}_d=0.99$). In this non-degenerative case, the SNr neurons fire very actively with $\langle f_i^{({\rm SNr})} \rangle = 25.5$ Hz.
Due to strong inhibitory projection from the SNr, the thalamic cells become silent, resulting in no movement (i.e., the BG door to the thalamus is locked in
the normal tonic default state).

But, with decreasing $x_{\rm D2}$ from 1 (degenerative case), as shown in Fig.~\ref{fig:Tonic}(h), ${\cal S}_{IP}$ is rapidly decreased from 23.4 to 1.4 , while
there is no change in ${\cal S}_{DP}$ (= 23.1). In this way, IP for the HD becomes weakened. Thus, as $x_{\rm D2}$ is decreased from 1, the competition degree ${\cal C}_d$ between DP and IP is found to increase from 0.99 to 16.5 [see Fig.~\ref{fig:Tonic}(i)]. Thus, balance between DP and IP becomes broken up in the degenerative tonic case.

Figures \ref{fig:Tonic}(j1)-\ref{fig:Tonic}(j4) show raster plots of spikes and IPSRs $R_{\rm SNr}(t)$ of the (output) SNr neurons for
$x_{\rm D2}$ = 1.0, 0.8, 0.5, and 0.2, respectively. We note that, firing activity of the SNr neurons becomes reduced with decreasing $x_{\rm D2}$
because of weakened IP. As a result of decrease in ${\cal S}_{IP}$ (strength of IP), the population-averaged MFR $\langle f_i^{\rm (SNr) } \rangle$ is found to decrease from 25.5 to 6.9 Hz with decreasing $x_{\rm D2}$ from 1, as shown in Fig.~\ref{fig:Tonic}(k). Thus, the BG gate to the thalamus becomes opened even in the case of tonic cortical input (3 Hz) in the resting state via break-up of balance between DP and IP.
Consequently, a tonic pathological state with involuntary jerky movement occurs, in contrast to the tonic default state without movement.

\begin{figure*}
\includegraphics[width=1.4\columnwidth]{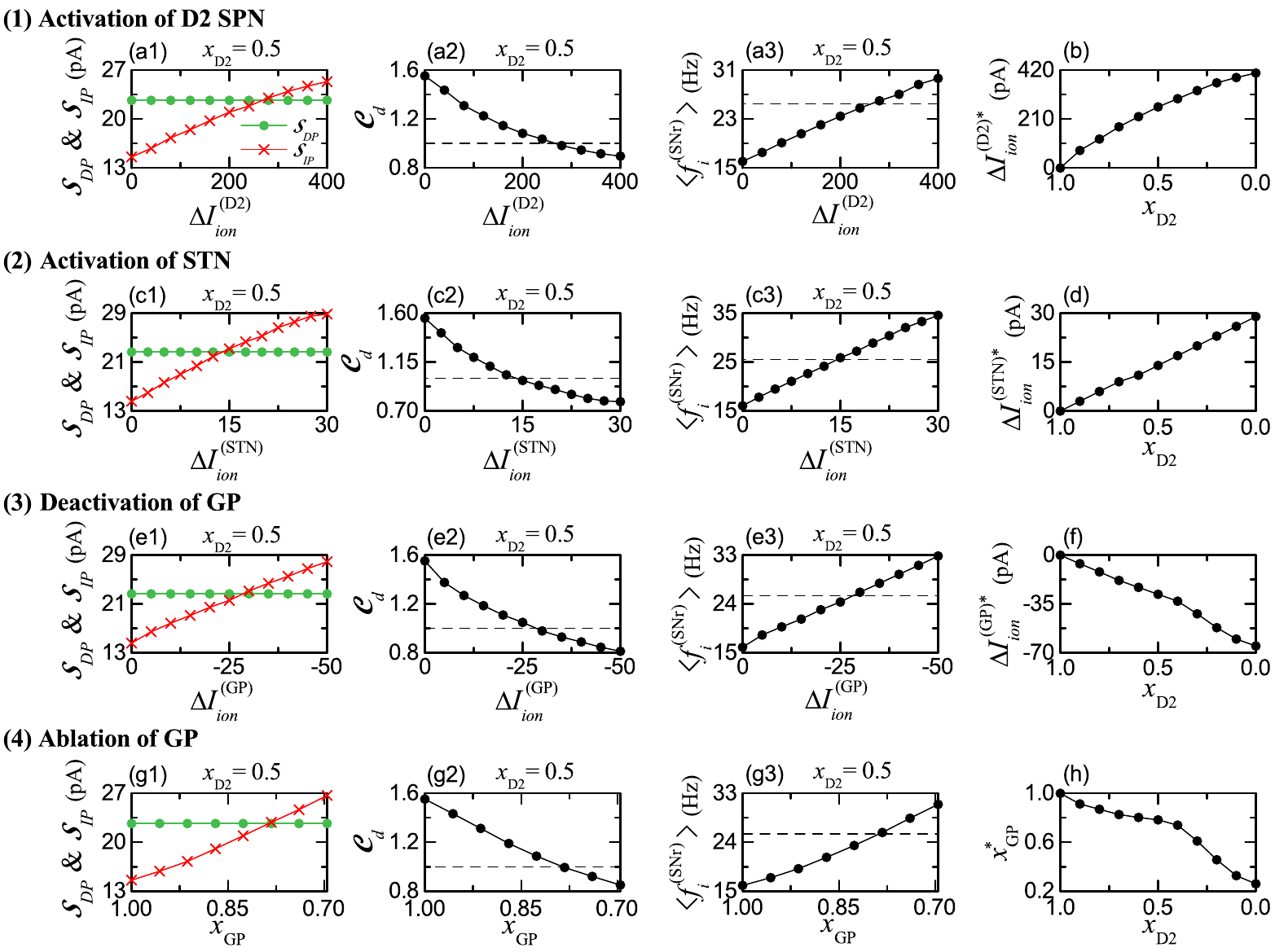}
\caption{Treatment of tonic pathological state by strengthening IP for the tonic cortical input (3 Hz). Colors: parts, related to DP (green), while parts, associated with IP (red). (1) Activation of D2 SPN: Plots of (a1) ${\cal{S}}_{DP}$ (green circles) and ${\cal{S}}_{IP}$ (red crosses), (a2) ${\cal{C}}_d$, and (a3) $\langle f_i^{({\rm SNr})} \rangle$ versus $\Delta I_{ion}^{({\rm D2})}$ for $x_{\rm D2}=0.5$. (b) Plot of threshold $\Delta I_{ion}^{({\rm D2})*}$ versus $x_{\rm D2}$. (2) Activation of STN: Plots of (c1) ${\cal{S}}_{DP}$ and ${\cal{S}}_{IP}$, (c2) ${\cal{C}}_d$, and (c3) $\langle f_i^{({\rm SNr})} \rangle$  versus $\Delta I_{ion}^{({\rm STN})}$ for $x_{\rm D2}=0.5$.
(d) Plot of threshold $\Delta I_{ion}^{({\rm D2})*}$ versus $x_{\rm D2}$. (3) Deactivation of GP: Plots of (e1) ${\cal{S}}_{DP}$  and ${\cal{S}}_{IP}$, (e2) ${\cal{C}}_d$, and (e3) $\langle f_i^{({\rm SNr})} \rangle$ versus $\Delta I_{ion}^{({\rm GP})}$ for $x_{\rm D2}=0.5$. (f) Plot of threshold $\Delta I_{ion}^{({\rm GP})*}$ versus $x_{\rm D2}$. (4) Ablation of GP: Plots of (g1) ${\cal{S}}_{DP}$ and ${\cal{S}}_{IP}$, (g2) ${\cal{C}}_d$, and (g3) $\langle f_i^{({\rm SNr})} \rangle$  versus $x_{\rm GP}$ for $x_{\rm D2}=0.5$. (h) Plot of threshold $x_{\rm GP}^*$ versus $x_{\rm D2}$. Horizontal dashed lines in (a2), (c2), (e2), and (g2) represent ${\cal{C}}_d^*$ (= 1.0) for the default healthy tonic state when $x_{\rm D2}=1$. Horizontal dashed lines in (a3), (c3), (e3), and (g3) represent $\langle f_i^{({\rm SNr})} \rangle$ (=25.5 Hz) for the default healthy tonic state when $x_{\rm D2}=1$.
}
\label{fig:TTreat}
\end{figure*}

Next, we consider the case of phasic cortical input (10 Hz) in the phasically active state \cite{Hump1,CI1,CI2,CI3,CI4,CI5,Str2,CN6,CN14}, which is shown in Fig.~\ref{fig:Phasic}. Population firing behavior of D1 SPNs, associated with DP (green color), is shown in their raster plot of spikes and the IPSR $R_{\rm D1}$(t) in Fig.~\ref{fig:Phasic}(a). In comparison to the tonic case with the population-averaged MFR $\langle f_i^{\rm (D1) } \rangle$ = 1.03 Hz in Fig.~\ref{fig:Tonic}(a), firing activity of the D1 SPNs become very active with $\langle f_i^{\rm (D1) } \rangle$ = 30.7 Hz, independently of $x_{\rm D2}$.

But, due to degenerative loss of D2 SPNs, population firing activities of the D2 SPNs, the STN neurons, and the GP neurons [related to the IP (red color)] are changed with decreasing $x_{\rm D2}$, as shown in their raster plots of spikes and IPSRs in Fig.~\ref{fig:Phasic}.
The population-averaged MFRs of the D2 SPNs, the STN neurons, and the GP neurons are also shown in Figs.~\ref{fig:Phasic}(c), \ref{fig:Phasic}(f), and \ref{fig:Phasic}(g), respectively. For the D2 SPNs, $\langle f_i^{\rm (D2) } \rangle$ = 24.1 Hz [much larger than that (0.97 Hz) in the tonic case], independently of $x_{\rm D2}$, because there is no change in cortical input to the D2 SPNs. As a result of decreased inhibitory projection from the D2 SPNs, $\langle f_i^{({\rm GP})} \rangle$ of the GP neurons is rapidly increased from 7.3 to 66.1 Hz with decreasing $x_{\rm D2}$ from 1; the increasing rate is higher than the tonic case.
On the other hand, due to increase in inhibitory projection from the GP, $\langle f_i^{({\rm STN})} \rangle$ of the STN neurons decreases from 39.8 to 17.6 Hz; the decreasing rate is also larger than that in the tonic case.

We consider the case of $x_{\rm D2}=1$ without degeneration. In this non-degenerative case, ${\cal S}_{DP}$ = 2309.7 and ${\cal S}_{IP}$ = 815.6.
Thus, the competition degree becomes ${\cal C}_d=2.82$ [i.e., ${\cal S}_{DP}$ (strength of DP) is 2.82 times larger than ${\cal S}_{IP}$ (strength of IP)].
In this case, $\langle f_i^{({\rm SNr})} \rangle$ of the (output) SNr neurons are decreased to 5.5 Hz (cf., in the tonic case, 25.5 Hz).
Consequently, the BG door to the thalamus becomes opened, leading to normal movement.
This phasic healthy state with ${\cal C}_d=2.82$ is in contrast to the tonic healthy state with ${\cal C}_d \simeq 1.0$ resulting in no movement.

However, as $x_{\rm D2}$ is decreased from 1 (degenerative case), ${\cal S}_{IP}$ is rapidly decreased from 815.6 to 92.3 , while
there is no change in ${\cal S}_{DP}$ (= 2309.7) [see Fig.~\ref{fig:Phasic}(h)]. Thus, IP becomes rapidly weakened.
Due to such under-activity of IP, the competition degree ${\cal C}_d$ increases from 2.82 (healthy state) to 25.0, as shown in Fig.~\ref{fig:Phasic}(i).
Consequently, harmony between DP and IP becomes broken up in the degenerative case with $x_{\rm D2} < 1$, and then a phasic pathological state with abnormal hyperkinetic movement appears, in contrast to the phasic healthy state with normal movement.

Raster plots of spikes and IPSRs $R_{\rm SNr}(t)$ of the (output) SNr neurons for $x_{\rm D2}$ = 1.0, 0.8, 0.5, and 0.2 are shown in
Figs.~\ref{fig:Phasic}(j1)-\ref{fig:Phasic}(j4), respectively. Due to under-activity of IP, firing activity of the SNr neurons becomes decreased with decreasing $x_{\rm D2}$ from 1.  Due to decreased ${\cal S}_{IP}$ (strength of IP), the population-averaged MFR $\langle f_i^{\rm (SNr) } \rangle$ decreases from 5.5 (healthy state) to 0.7 Hz with decreasing $x_{\rm D2}$ from 1 [see Fig.~\ref{fig:Phasic}(k)]. In this phasic pathological state with ${\cal C}_d > 2.82$ (where harmony between DP and IP is broken up), abnormal hyperkinetic movement disorder occurs, in contrast to the normal movement
for the phasic healthy state with ${\cal C}_d = 2.82$ (where there is harmony between DP and IP).

To sum up the above results briefly, it is shown that, for the HD, pathological states (where harmony between DP and IP is broken up) appear due to degenerative loss of D2 SPNs in the cases of both tonic and phasic cortical inputs. On the other hand, for the PD, pathological states appear because of DA deficiency
\cite{PD1,PD2,PD3,PD4,PD5,PD6,KimPD}. In the case of HD, IP is under-active, in contrast to the case of PD with over-active IP.
Thus, patients with HD exhibit abnormal hyperkinetic movement disorder, while patients with PD show abnormal hypokinetic movement disorder.
Consequently, HD lies at one end of the spectrum of movement disorders in the BG, while PD lies at the other end.

\subsection{Treatment of HD via Recovery of Harmony between DP and IP}
\label{subsec:Treat}
For the pathological state in the HD, IP is under-active due to degenerative loss of D2 SPNs, in comparison to the healthy state.
Thus, harmony between DP and IP is broken up (i.e. occurrence of disharmony between DP and IP), leading to abnormal hyperkinetic movement disorder.
Here, based on optogenetics \cite{OG1,OG2}, we investigate treatment of the pathological state with enhanced competition degree ${\cal C}_d$ (than the normal one for the healthy state) in both cases of tonic and phasic cortical inputs via recovery of harmony between DP and IP.

Optogenetics is a control technique for the activity of target cells in living organisms by combining optics and genetics. The target cells are genetically modified to express opsins (light-sensitive proteins) (i.e., fusion of the opsins into the target cells). When the opsins are activated by the light stimulation with
specific wavelengths, variation in the intrinsic ionic currents of the cells in the target population $X$, $\Delta I_{ion}^{(X)}$, takes place
\cite{OG1,OG2}. If $\Delta I_{ion}^{(X)}$ is positive (negative), firing activity of the target cells is increased (decreased), resulting in their activation (deactivation). Such activation and deactivation of the target cells was studied in our recent work \cite{KimPD}. As discussed in \cite{KimPD}, we simulate the effect of optogenetics by adding $\Delta I_{ion}^{(X)}$ in Eq. (A1) in Appendix A in \cite{KimPD}, in addition to the current, $I_i^{(X)}$, into the target $X$ population. With increasing the intensity of light stimulation, the magnitude of $\Delta I_{ion}^{(X)}$ also increases.

We first consider tonic pathological states with enhanced competition degree ${\cal C}_d$ [larger than that (1) for the tonic healthy state (with balanced DP and IP)], occurring due to degenerative loss of D2 SPNs, in the case of tonic cortical input (3 Hz) (see Fig.~\ref{fig:Tonic}). As an example, we consider the tonic pathological case of $x_{\rm D2}=0.5$ with ${\cal C}_d=1.53$. In this pathological case, IP is under-active in comparison to the tonic healthy case (with balanced DP and IP); firing activity of D2 SPNs is under-active, leading to over-activity of GP neurons, which then results in under-activity of the STN neurons. Hence, for recovery of balance between DP and IP, we try to strengthen the IP via activation of D2 SPNs and STN neurons and deactivation of GP neurons.

We first strengthen the IP through activation of the target (under-active) D2 SPNs.
Figure \ref{fig:TTreat}(a1) shows plots of ${\cal S}_{IP}$ (strength of IP) and ${\cal S}_{DP}$ (strength of DP) versus $\Delta I_{ion}^{({\rm D2})}$.
As $\Delta I_{ion}^{({\rm D2})}$ is increased from 0,  ${\cal S}_{IP}$ (red) increases from 15.1, while ${\cal S}_{DP}$ (green) remains unchanged (i.e., 23.1). As a result of increase in ${\cal S}_{IP}$, the competition degree ${\cal C}_d$ between DP and IP is found to decrease from 1.53 [Fig.~\ref{fig:TTreat}(a2)]. Also, the population-averaged MFR of the output SNr neurons, $\langle f_i^{({\rm SNr})} \rangle$, is found to increase from 16.1 Hz [Fig.~\ref{fig:TTreat}(a3)].

We note that, as $\Delta I_{ion}^{({\rm D2})}$ passes a threshold $\Delta I_{ion}^{({\rm D2})*}$ (= 262 pA), ${\cal C}_d~=~ {\cal C}_d^*~(=~1.0)$ and
$\langle f_i^{({\rm SNr})} \rangle~ =~ \langle f_i^{({\rm SNr})*} \rangle$ (= 25.5 Hz); ${\cal C}_d^*$ and $\langle f_i^{({\rm SNr})*} \rangle$ are those for the tonic healthy state, and they are represented by the horizontal dashed lines in Figs.~\ref{fig:TTreat}(a2) and \ref{fig:TTreat}(a3).
Thus, for $x_{\rm D2}=0.5$, the pathological state with ${\cal C}_d = 1.53$ may have ${\cal C}_d$ (= 1.0) via activation of D2 SPNs for
the threshold, $\Delta I_{ion}^{({\rm D2})*}$ (= 262 pA); DP and IP becomes balanced, as in the case of tonic healthy state.
In this way, balance between DP and IP is recovered for $\Delta I_{ion}^{({\rm D1})*}$ = 262 pA.
Figure \ref{fig:TTreat}(b) shows the plot of $\Delta I_{ion}^{({\rm D2})*}$ versus $x_{\rm D2}$.
As $x_{\rm D2}$ is decreased from 1, the threshold $\Delta I_{ion}^{({\rm D2})*}$ is increased;
with decreasing $x_{\rm D2}$, more $\Delta I_{ion}^{({\rm D2})*}$ is necessary for recovery of balance between DP and IP.

We also strengthen the IP via activation of the target (under-active) STN neurons, which is shown in Figs.~\ref{fig:TTreat}(c1)-\ref{fig:TTreat}(c3) for $x_{\rm D2}=0.5$. All the behaviors are qualitatively the same as those in the case of activation of D2 SPNs.
With increasing $\Delta I_{ion}^{({\rm STN})}$ from 0, ${\cal S}_{IP}$ (strength of IP) increases, leading to decrease in the competition degree
${\cal C}_d$, and the population-averaged MFR of the output SNr neurons, $\langle f_i^{({\rm SNr})} \rangle$, also increases.
But, the threshold $\Delta I_{ion}^{({\rm STN})*}$ (= 14 pA), where balance between DP and IP is recovered (i.e., ${\cal C}_d$ = 1 and
$\langle f_i^{({\rm SNr})} \rangle$ = 25.5 Hz, as in the case of tonic healthy state), is smaller than that (262 pA) in the case of activation of D2 SPNs.
The mono-synaptic effect of STN neurons on the output SNr neurons is more direct than the bi- or tri-synaptic effect of D2 SPNs, which could result in the smaller
threshold $\Delta I_{ion}^{({\rm STN})*}$  in the case of STN neurons.
Figure \ref{fig:TTreat}(d) shows the plot of $\Delta I_{ion}^{({\rm STN})*}$ versus $x_{\rm D2}$.
With decreasing $x_{\rm D2}$ from 1, the threshold $\Delta I_{ion}^{({\rm STN})*}$ increases, as shown in Fig.~\ref{fig:TTreat}(d);
As $x_{\rm D2}$ is decreased, more $\Delta I_{ion}^{({\rm STN})*}$ is necessary for recovery of balance between DP and IP.

\begin{figure*}
\includegraphics[width=1.4\columnwidth]{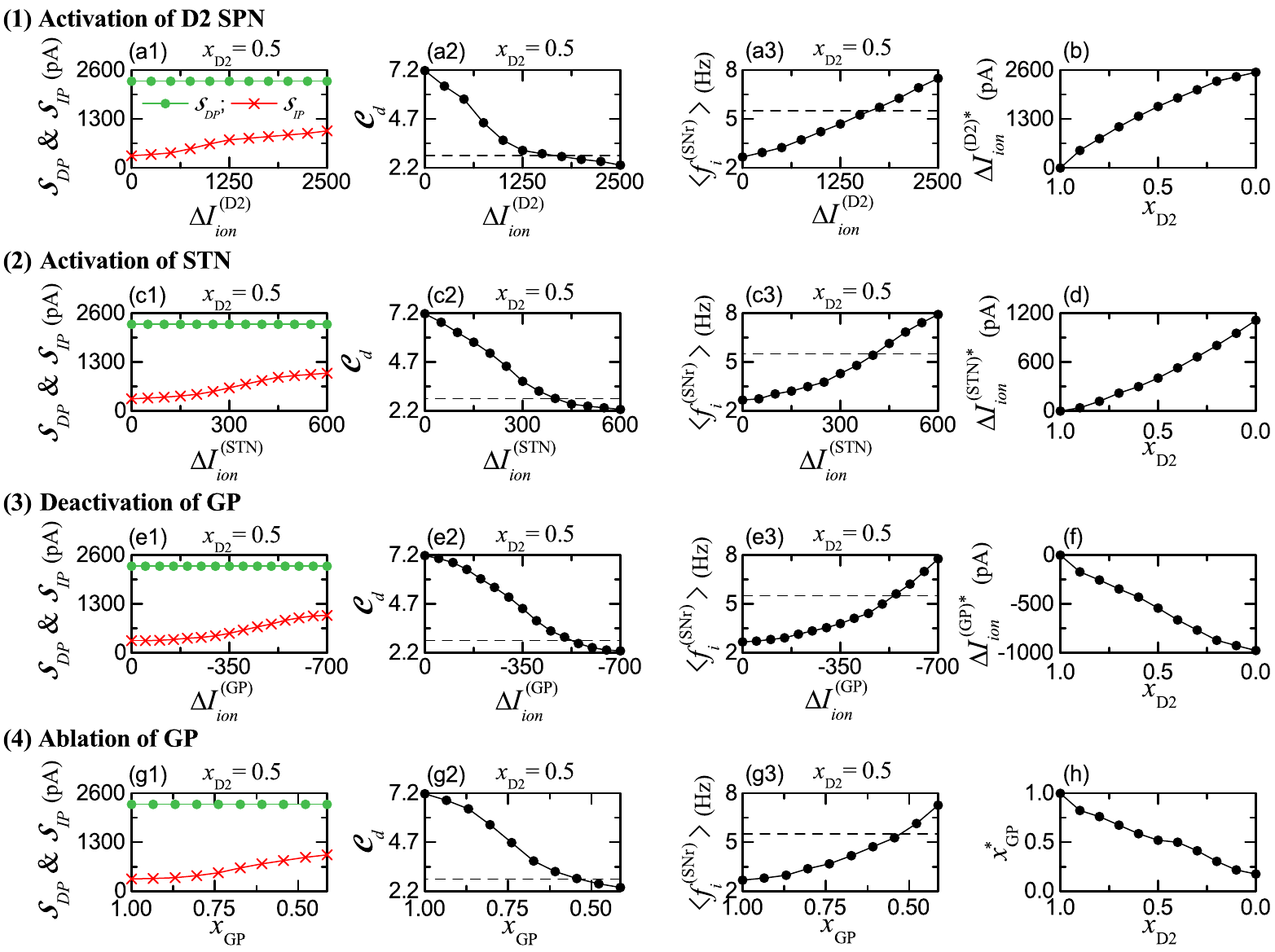}
\caption{Treatment of phasic pathological state by strengthening IP for the phasic cortical input (10 Hz).
Colors: parts, related to DP (green), while parts, associated with IP (red).
(1) Activation of D2 SPN: Plots of (a1) ${\cal{S}}_{DP}$ and ${\cal{S}}_{IP}$, (a2) ${\cal{C}}_d$, and (a3) $\langle f_i^{({\rm SNr})} \rangle$ versus $\Delta I_{ion}^{({\rm D2})}$ for $x_{\rm D2}=0.5$. (b) Plot of threshold $\Delta I_{ion}^{({\rm D2})*}$ versus $x_{\rm D2}$. (2) Activation of STN: Plots of (c1) ${\cal{S}}_{DP}$ and ${\cal{S}}_{IP}$, (c2) ${\cal{C}}_d$, and (c3) $\langle f_i^{({\rm SNr})} \rangle$ versus $\Delta I_{ion}^{({\rm STN})}$ for $x_{\rm D2}=0.5$. (d) Plot of threshold $\Delta I_{ion}^{({\rm STN})*}$ versus $x_{\rm D2}$. (3) Deactivation of GP: Plots of (e1) ${\cal{S}}_{DP}$ and ${\cal{S}}_{IP}$, (e2) ${\cal{C}}_d$, and (e3) $\langle f_i^{({\rm SNr})} \rangle$ versus $\Delta I_{ion}^{({\rm GP})}$ for $x_{\rm D2}=0.5$. (f) Plot of threshold $\Delta I_{ion}^{({\rm GP})*}$ versus $x_{\rm D2}$. (4) Ablation of GP: Plots of (g1) ${\cal{S}}_{DP}$ and ${\cal{S}}_{IP}$, (g2) ${\cal{C}}_d$, and (g3) $\langle f_i^{({\rm SNr})} \rangle$ versus $x_{\rm GP}$ for $x_{\rm D2}=0.5$. (h) Plot of threshold $x_{\rm GP}^*$ versus $x_{\rm D2}$. Horizontal dashed lines in (a2), (c2), (e2), and (g2) represent ${\cal{C}}_d^*$ (= 2.82) for the healthy phasic state when $x_{\rm D2}=1$. Horizontal dashed lines in (a3), (c3), (e3), and (g3) represent $\langle f_i^{({\rm SNr})} \rangle$ (= 5.5 Hz) for the healthy phasic state when $x_{\rm D2}=1$.
}
\label{fig:PTreat}
\end{figure*}

Unlike the cases of activation of (under-active) D2 SPNs and STN neurons, IP may be strengthened via deactivation of (over-active) GP neurons;
in the case of deactivation, $\Delta I_{ion}^{(GP)}$ is negative, in contrast to the case of activation with $\Delta I_{ion}^{(X)} > 0$ [$X$ = D2 (SPN) and STN].
Figures \ref{fig:TTreat}(e1)- \ref{fig:TTreat}(e3) and \ref{fig:TTreat}(f) show the case of deactivation of GP neurons.
As the magnitude of $\Delta I_{ion}^{({\rm GP})}$ is increased (i.e., more negative), strength of IP, ${\cal S}_{IP}$ (red), is found to increase from 15.1, while
${\cal S}_{DP}$ (green) remains constant (= 23.1). Thus, when passing a threshold $\Delta I_{ion}^{({\rm GP})*} = -28$ pA, balance between DP and IP becomes recovered (i.e., the competition degree ${\cal C}_d$ becomes 1 and the population-averaged MFR of output SNr neurons $\langle f_i^{({\rm SNr})} \rangle$ becomes 25.5 Hz) [see Figs.~\ref{fig:TTreat}(e2)- \ref{fig:TTreat}(e3)]. As shown in Fig.~\ref{fig:TTreat}(f),
with decreasing $x_{\rm D2}$ from 1, the threshold $\Delta I_{ion}^{({\rm GP})*}$ is decreased (i.e., its magnitude increases);
as $x_{\rm D2}$ is decreased from 1, more negative $\Delta I_{ion}^{({\rm GP})*}$ is required for recovery of balance between DP and IP.

Instead of the above deactivation of GP neurons via optogenetics, we also consider ablation of (over-active) GP neurons in the pathological state for $x_{\rm D2}= 0.5$ to reduce the over-activity of GP neurons. In the case of ablation, the number of GP neurons, $N_{{\rm GP}}$, is reduced to $N_{\rm GP}^{(n)}$ $x_{\rm GP}$
($ 1 > x_{\rm GP} > 0)$, where $N_{\rm GP}^{(n)}$ (= 46) is the normal number of GP neurons and $x_{\rm GP}$ is the fraction of number of GP neurons.
As shown in Figs.~\ref{fig:TTreat}(g1)- \ref{fig:TTreat}(g3) and \ref{fig:TTreat}(h), the effect of decreasing $x_{\rm GP}$ via ablation is similar to that of deactivation of GP neurons via optogenetics. As $x_{\rm GP}$ is decreased from 1, strength of IP, ${\cal S}_{IP}$ (red), is found to increase from 15.1 (i.e., IP becomes strengthened) [see Fig.~\ref{fig:TTreat}(g1)]. When passing a threshold, $x_{\rm GP}^*~(\simeq 0.78)$, balance between DP and IP becomes recovered
(i.e., ${\cal C}_d$ = 1.0 and $\langle f_i^{({\rm SNr})} \rangle$ = 25.5 Hz), as shown in Figs.~\ref{fig:TTreat}(g2)-\ref{fig:TTreat}(g3).
Figure \ref{fig:TTreat}(h) shows the plot of $x_{\rm GP}^*$ versus $x_{\rm D2}$.
With decreasing $x_{\rm D2}$ from 1, $x_{\rm GP}^*$ decreases; more ablation (i.e., smaller $x_{\rm GP}$) is necessary for balance between DP and IP.

Next, we consider phasic pathological states with enhanced competition degree ${\cal C}_d$ [larger than that (2.82) for the phasic healthy state (with harmony between DP and IP)], occurring due to degenerative loss of D2 SPNs, in the case of phasic cortical input (10 Hz) (see Fig.~\ref{fig:Phasic}). As an example, we consider the pathological case of $x_{\rm D2}=0.5$ with ${\cal C}_d=7.19$. In this phasic pathological case, IP is under-active in comparison to the case of phasic healthy state. For the phasic healthy state with ${\cal C}_d^*$ = 2.82 (i.e., harmony between DP and IP), the population-averaged MFR of output STr neurons, $\langle f_i^{({\rm SNr})*} \rangle,$ is much reduced to 5.5 Hz, leading to normal movement, in contrast to the case of tonic healthy state with ${\cal C}_d \simeq 1.0$ and $\langle f_i^{({\rm SNr})} \rangle$ = 25.5 Hz without movement. As in the above tonic pathological state, firing activity of D2 SPNs is under-active, resulting in over-activity of GP neurons, which then leads to  under-activity of the STN neurons. Hence, for recovery of harmony between DP and IP, we strengthen the IP through activation of D2 SPNs and STN neurons and deactivation of GP neurons by employing optogenetic technique and via ablation of GP neurons.

Figure \ref{fig:PTreat} shows treatment of phasic pathological state for $x_{\rm D2}=0.5$ with ${\cal C}_d=7.19$; (1) activation of D2 SPNs, (2) activation of STN neurons, (3) deactivation of GP neurons, and (4) ablation of GP neurons. The overall results of these treatments are qualitatively the same as those in the above case of tonic pathological state in Fig.~\ref{fig:TTreat}. Only the corresponding thresholds are quantitatively different;
(1) $\Delta I_{ion}^{({\rm D2})*}$ = 1,636 pA, (2) $\Delta I_{ion}^{({\rm STN})*}$ = 405 pA, (3) $\Delta I_{ion}^{({\rm GP})*} = - 540$ pA, and
(4) $x_{\rm GP}^*~(\simeq 0.52)$. When passing a threshold for each treatment, harmony between DP and IP becomes recovered (i.e.,
${\cal C}_d$ = 2.82 and $\langle f_i^{({\rm SNr})} \rangle$ = 5.5 Hz), resulting in normal movement.
Finally, we note that, with decreasing $x_{\rm D2}$, the thresholds, $\Delta I_{ion}^{({\rm D2})*}$ and $\Delta I_{ion}^{({\rm STN})*}$, for activations of D2 SPNs and STN neurons are increased (i.e., more positive), and the threshold, $\Delta I_{ion}^{({\rm GP})*}$ for deactivation of GP neurons becomes more negative, as shown in Figs.~\ref{fig:PTreat}(b), \ref{fig:PTreat}(d), and \ref{fig:PTreat}(f). Thus, as $x_{\rm D2}$ is decreased, more light stimulation for activation and deactivation is necessary for recovery of harmony between DP and IP. Also, in the case of ablation of GP neurons, with decreasing $x_{\rm D2}$, more ablation
is required to get harmony between DP and IP.

\begin{figure}
\includegraphics[width=1.0\columnwidth]{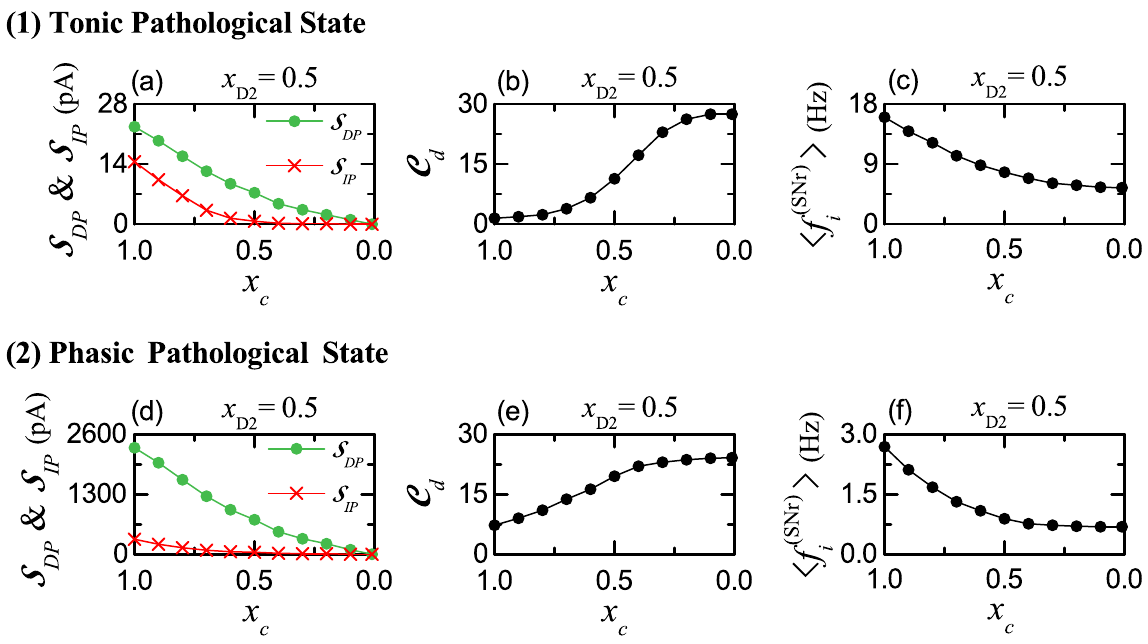}
\caption{Effect of loss of healthy synapses in all the BG neurons on the HD for $x_{\rm D2}=0.5$. Colors: parts, related to DP (green), while parts, associated with IP (red). (1) Tonic pathological state: Plots of (a) ${\cal{S}}_{DP}$ and ${\cal{S}}_{IP}$, (b) ${\cal{C}}_d$, and (c) $\langle f_i^{({\rm SNr})} \rangle$ versus $x_c$. (2) Phasic pathological state: Plots of (d) ${\cal{S}}_{DP}$  and ${\cal{S}}_{IP}$, (e) ${\cal{C}}_d$, and (f) $\langle f_i^{({\rm SNr})} \rangle$ versus $x_c$.
}
\label{fig:SL}
\end{figure}

\subsection{Effect of Loss of Healthy Synapses on The HD}
\label{subsec:SynEffect}
In the HD, loss of healthy synapses occurs not only in the striatum, but also in other regions of the BG, including STN, GP, and SNr \cite{Loss1,Loss2,Loss3,Loss4,Loss5,Loss6}. Such loss of synapses in the BG neurons is an important feature of HD, and it is thought to contribute to the motor and cognitive symptoms of the disease. Here, we study effect of loss of healthy synapses of all the BG neurons on HD.

As examples, we consider pathological states for $x_{\rm D2}$ = 0.5 in both cases of tonic (3 Hz) and phasic (10 Hz) cortical inputs.
Loss of synapses in the BG neurons is modeled in terms of decreased synaptic connection probability,
$p_c = p_c^{(n)}~x_c$; $p_c^{(n)}$ is the normal synaptic connection probability (depending on the type of BG neurons and given
in Table II in \citep{KimPD}) and $x_c$ represents the fraction in $p_c$ ($1 > x_c > 0$).

We first consider a tonic pathological state in Figs.~\ref{fig:SL}(a)-\ref{fig:SL}(c).
As a result of loss of synapses, decreased cortical inputs into D1 SPNs leads to reduction in their firing activity $\langle f_i^{({\rm D1})} \rangle$.
Then, strength of DP, ${\cal S}_{DP}$, becomes decreased. As shown in Fig.~\ref{fig:SL}(a), ${\cal S}_{DP}$ (green color) is found to monotonically decrease from 23.1 with decreasing $x_c$ from 1. Also, due to reduced cortical synaptic inputs into D2 SPNs, firing activity of D2 SPNs, $\langle f_i^{({\rm D2})} \rangle$, becomes decreased, leading to increased firing activity of GP neurons ($\langle f_i^{({\rm GP})} \rangle$), which then results in decrease in the firing activity of STN neurons ($\langle f_i^{({\rm STN})} \rangle$). Consequently, strength of IP, ${\cal S}_{IP}$, becomes decreased.
In this case of IP, with decreasing $x_c$ from 1, ${\cal S}_{IP}$ (red color) is found to more rapidly decrease from 15.1 than the case of DP.
Then, the competition ${\cal C}_d$ between DP and IP increases rapidly from 1.53 with decreasing $x_c$ from 1 [see Fig.~\ref{fig:SL}(b)].
Thus, as $x_c$ is decreased from 1, population-averaged MFR of the output SNr neurons ($\langle f_i^{({\rm SNr})} \rangle$) decreases from 16.1 Hz.
In this way, with decreasing $x_c$, the degree of disharmony between DP and IP becomes increased, resulting in more severe involuntary jerky movement in the tonic pathological case.

Next, we consider a phasic pathological state in Figs.~\ref{fig:SL}(d)-\ref{fig:SL}(f).
With decreasing $x_c$, tendency in the phasic pathological case is qualitatively the same as that in the above tonic pathological case.
Based on the same reasoning given in the tonic pathological case, as $x_c$ is decreased from 1, strength of IP (${\cal S}_{IP}$; red color) is found to decreases much more rapidly than strength of DP (${\cal S}_{DP}$; green color), as shown in Fig.~\ref{fig:PTreat}(d).
Then, the competition degree ${\cal C}_d$ increases from 7.19 with decreasing $x_c$ from 1 [see Fig.~\ref{fig:PTreat}(e)].
Consequently, firing activity of the output SNr neurons ($\langle f_i^{({\rm SNr})} \rangle$) decreases from 2.7 Hz with $x_c$, as shown in Fig.~\ref{fig:PTreat}(f).
In this way, as $x_c$ is decreased from 1, the broken-up degree of harmony between DP and IP becomes increased, leading to more severe abnormal hyperkinetic movement disorder in the phasic pathological case.

Overall, in both tonic and phasic pathological cases, as a result of loss of healthy synapses in the BG neurons, symptoms of the HD become more severe
with decreasing $x_c$, because the disharmony degree between DP and IP becomes increased.

\section{Summary and Discussion}
\label{sec:SUM}
The BG exhibit a variety of functions for motor, cognition, and emotion. Dysfunction in the BG is associated with movement disorder (e.g., HD and PD) and cognitive and psychiatric disorders. There are two competing pathways in the BG, DP (facilitating movement) and IP (suppressing movement) \cite{DIP1,DIP2,DIP3,DIP4}. In our recent work \cite{KimPD}, as a first time, we made quantitative analysis of competitive harmony between DP and IP in the default tonic state and the phasic healthy and pathological states by introducing their competition degree, ${\cal C}_d,$ between DP and IP, given by the ratio of strength of DP (${\cal S}_{DP})$ to strength of IP (${\cal S}_{IP})$ (i.e., ${\cal C}_d = {\cal S}_{DP} / {\cal S}_{IP}$).

In our prior work \cite{KimPD}, we studied PD which was found to occur for lower DA level. In the case of PD with reduced competition degree, IP is over-active, while DP is under-active, leading to abnormal hypokinetic movement.

In this paper, we are concerned in the HD which is a genetic neurodegenerative disease. We considered the early stage of HD where the DA level is normal.
As a result of mutant HTT gene, toxic HTT protein aggregates appear, causing the characteristic neurodegeneration seen in HD.
We considered degenerative loss of D2 SPNs in the case of normal DA level.
By decreasing $x_{\rm D2}$ (i.e. fraction of number of D2 SPNs) from 1, we quantitatively analyzed break-up of harmony between DP and IP.
IP was found to be under-active (i.e., weakened), in contrast to the case of PD with over-active IP. Thus, the competition degree ${\cal C}_d$  becomes increased than normal one. Consequently, abnormal hyperkinetic movement such as chorea occurs, in contrast to the case of PD with hypokinetic disorder.

Unfortunately, at present there is no cure for HD. The available treatments for HD primarily aim to control and alleviate its symptoms, resulting from weakened IP:
medication treatment \cite{Treat1,Treat2,Treat3,Treat4}, reducing symptoms, deep brain stimulation in research and clinical trials \cite{Treat5,Treat6,Treat7,CNYu1}, and experimental surgery \cite{Treat8}. Here, we studied treatment of HD via recovery of harmony between DP and IP by activating D2 SPNs and STN neurons and deactivating GP neurons, based on optogenetics \cite{OG1,OG2}. Through the treatment process, the IP becomes strengthened, and thus harmony between DP and IP may be regained. The results for the 3 optogenetic targets (D2 SPN, STN, GP) are well shown in Figs.~\ref{fig:TTreat}-\ref{fig:PTreat}. We note that in
the case of STN, the magnitude of threshold $\Delta I_{ion}^{\rm (STN)*}$ where harmony between DP and IP is recovered is the lowest, because the mono-synaptic effect of STN neurons on the output nucleus SNr is more direct than the bi- and tri-synaptic effect of D2 SPNs and GP cells. Hence, the STN could be the most effective target for optogenetic treatment. We also studied effects of loss of healthy synapses of the BG cells on the HD. As healthy synapses are lost in the BG, strength of the IP is found to decrease more rapidly than the case of the DP, resulting in increase in disharmony between DP and IP increases. Consequently, symptoms of the HD become worse. In this case of synapse loss, optogenetics and GP ablation are also expected to be effectively used for treatment. But, the threshold values of treatment could increase in comparison to the case without loss of healthy synapses, because the symptoms of HD are worse in the case of synapse loss.

We also make overall summary of our basic underlying approach for study of the BG function. The SNr is the output nucleus of the BG, providing inhibitory projection to the thalamus. Firing activity of the SNr is well characterized in terms of their population-averaged MFR $\langle f_i^{({\rm SNr})} \rangle$.
When $\langle f_i^{({\rm SNr})} \rangle$ is high (low), the BG gate to the thalamus becomes locked (opened), leading to inhibition (disinhition) of the thalamus. In this way, the population-averaged MFR of the SNr, $\langle f_i^{({\rm SNr})} \rangle$, is a good indicator for the functional activity of the BG. So, 
in Figs.~\ref{fig:Tonic}-\ref{fig:PTreat}, we examined variation in $\langle f_i^{({\rm SNr})} \rangle$ with respect to change in $x_{D2}$ (fraction of the number of D2 SPNs) and in variation in intrinsic ionic current $\Delta I_{ion}^{(X)}$ due to optogenetics. 

We note that firing activity (i.e. MFR) of the SNr is determined via competition between the DP synaptic current and the IP synaptic current into the SNr. Their competition may be well characterized in terms of our recently-introduced competition degree $C_d$, given by the ratio of the strength of DP to the strength of IP \cite{KimPD}. Thus, $C_d$ plays a good role of indicator for the synaptic inputs into the SNr, in contrast to the output indicator, $\langle f_i^{({\rm SNr})} \rangle$. Hence, relationship between $C_d$ and $\langle f_i^{({\rm SNr})} \rangle$ may be regarded as the cause-and-effect. The larger $C_d$ is, the lower
$\langle f_i^{({\rm SNr})} \rangle$. We obtained $C_d$ and $\langle f_i^{({\rm SNr})} \rangle$ in our BG SNN, well shown in Figs.~\ref{fig:Tonic}-\ref{fig:PTreat}. In this sense, we emphasize that $C_d$ and $\langle f_i^{({\rm SNr})} \rangle$ are basic quantities characterizing the BG functional activity.
In future, it would be interesting to try to get $C_d$ and $\langle f_i^{({\rm SNr})} \rangle$ in the mean field model for comparison \cite{MF}. For such comparison, mean-field approach must be developed to obtain not only the population-averaged output $\langle f_i^{({\rm SNr})} \rangle$ of the SNr, but also the population-averaged DP and IP synaptic input currents into the SNr.

Finally, we discuss limitations of our present work and future works.
In the present work, we considered early stage of HD where degenerative loss of D2 SPNs occurs in the nearly normal DA level.
But, in the late stage of HD, degenerative loss of D1 SPN also occurs along with decrease in DA level, leading to hypokinetic
disorder (e.g., rigidity and bradykinesia) due to weakened DP, as in the case of PD \cite{Dege-D1}.
Moreover, in addition to deaths of D1/D2 SPNs, degeneration of cortical pyramidal cells occurs \cite{Dege-C1,Dege-C2}.
Hence, as a future work, it would be interesting to investigate consequences of degeneration of D1 SPNs and cortical pyramidal cells, in addition to
degenerative loss of D2 SPNs.

Next, we would like to consider more realistic striatal circuit in the BG.
In our present striatal circuit, we considered only the D1/D2 SPNs (95 $\%$ major population).
But, the minor population of fast interneurons (FSIs) in the striatum are known to exert strong effects on firing activities of the D1/D2 SPNs \cite{Str2,FSI}.
Hence, in future, it would be worth while to contain the FSIs in our BG SNN.
In addition, lateral connections between D1 SPNs and D2 SPNs also exists in the striatum \cite{CN11,SPN2}. Thus, it would be worth while to contain lateral connections between D1 SPNs and D2 SPNs in our BG SNN and investigate the HD state and treatment.
In our present BG SNN, cortical inputs were modelled by Poisson spike trains. Such SNN could be extended to the cortico-BG-thalamo-cortical
(CBGTC) loop by including the cortical and the thalamic neurons for more complete computational work \cite{CN1,CBGTC}.

We also make discussion on application of the optogenetics to human patients for treatment of a pathological state \cite{OG3,OG4}.
In the case of HD, harmony between DP and IP is broken up due to under-active IP.
As shown in Sec.~\ref{subsec:Treat}, harmony between DP and IP could be recovered by strengthening IP. To this end, optogenetic techniques may be employed. Activation of D2 SPNs and STN neurons via optogenetics results in strengthening IP.
We hope that, in near future, safe clinical applications of optogenetics to human patients with HD could be successfully available
via collaboration of researchers and clinicians. Then, it would take a substantial step forward for treatment of HD.

We note that the optogenetic treatment could have benefits in comparison to the traditional deep-brain-stimulation (DBS) treatment. The DBS has the following disadvantages \cite{OG3,OG4}; (a) it is difficult to accurately determine the target cells, leading to cause many side effects and (b) a process with many trial and errors is necessary to each patient for optimal control. On the other hand, the target cells can be accurately located by optogenetic stimulation. Hence, side effects and trial-and-error process may be decreased in spite of limitations for application to the human patients \cite{OG3}.

\section*{Acknowledgments}
This research was supported by the Basic Science Research Program through the National Research Foundation of Korea (NRF) funded by the Ministry of Education (Grant No. 20162007688).

\end{document}